\documentclass[12pt]{article}
\usepackage{amssymb}
\usepackage{graphicx}
\usepackage{amsmath}

\def\vx{{\bf x}}

\newcommand{\bc}{\begin{center}}
\newcommand{\ec}{\end{center}}
\newcommand{\beq}{\begin{equation}}
\newcommand{\eeq}{\end{equation}}
\newcommand{\ba}{\begin{eqnarray*}}
\newcommand{\ea}{\end{eqnarray*}}

\newcommand{\ket}[1]{\ensuremath{\vert\,#1\,\rangle}}
\newcommand{\braket}[2]{\ensuremath{\langle\,#1\,\vert\,#2\,\rangle}}

\newcommand{\dee}[2]{\ensuremath{{{\rm d} #1\over {\rm d} #2}}}

\def\ann{{\hat{a}}}

\def\cre{{\hat{a}^+}}

\def\Lio{{\hat{L}}}

\begin{document}
\title {Field theoretic approach in kinetic reaction:
 Role of random sources and sinks }
 \maketitle
\author{M. Hnatich$^{1,2}$, J. Honkonen$^{3}$, T. Lucivjansky$^{1,2}$}
%
 $^{1}$ Institute of Experimental
Physics, Slovak Academy of Sciences, Watsonova 47, 040 01
Ko\v{s}ice, Slovakia,\\ 
$^{2}$ Faculty of Sciences, P.J. \v{S}afarik
University, Moyzesova 16, 040 01 Ko\v{s}ice, Slovakia,\\ 
$^3$
Department of Military Technology, National Defence University,
P.O.~Box~7, 00861, Helsinki, Finland.\\

\begin{abstract}
The effects  of random sources  and sinks on the reaction kinetics 
have been investigated in the framework of field theoretic model, which is the result of Doi 
"second quantization" of corresponding master equation . It has been  
demonstrated that they  do not affect strong impact of  density fluctuations on
the single-species annihilation reaction  in two dimension. However random source nad sinks  
generate some new terms which have to be added into equation describing Gribov process. 
\end{abstract}


\section{Introduction}
\label{sec:intro}
Reaction kinetics - main subject of our analysis belongs to the spreading processes, 
which attract much attention in last time due to their frequent occurrance in physical, 
chemical, biological, ecological and sociological systems. 
Generally, such systems are characterised by large fluctuations of physical quatities 
therefore it must be described by the methods of statistical physics. 
At the same time appropriate quantities, which give full information about system, 
must be introduced, e.g. probability distribution functions.  For systems with enormously 
large number of particles (e.g. gas, plasma etc.) they satisfy known chain of evolution equations. 
Reaction kinetics, which can be afilled to the system with large number of particles,   
operates with processes, where the number of particles is not conserved and    probability 
distibution functions for fluctuating random quantities satisfy the master equations. 
Due to elegant and robust technique \cite{Doi76} these equations can be rewritten in 
the terminology of a "quantum theory" and after that by means of technique of functional 
integration can be formulated as  the effective field-theoretic models with a given action and defined random fields.
These models describe mumerous processes with the same universality classes. 
It suggests situation with  various field- theoretic models of critical dynamics obtained 
from Langevin equation\cite{Vasiljev}. The Langevin equation is a stochastic equation with 
random forcing, which enters to the equation as additive or multiplicative factor, compensates dissipative losses and sinks
 and, hence, ensures 
steady state of dynamics under consideration.  
In reaction kinetics the random sources and sinks, in fact,  reflect real physical situation, 
in which during chemical reaction of follow up species, they can appear or disapear due to 
uncontroled random interaction with bath. , e.g. due to active chemical radicals. 
Theoretical description of such a sources and sinks  is not easy task. 
Their direct inclusion as defined additive regular terms and noises  into 
Langevin equation is artificial and their physical interpretation is dubious. 
There exists another physically more acceptable way,namely, introduction of new terms,
responsible for interactions with bath,  into corresponding master equation.
In present paper we propose specific realization of general approach eleborated in \cite{vanKampen84} and we 
add the random sources and sinks to the master equation for annihilation equation. At the beginning 
we introduce the simplest terms responsible  for the sources and sinks. 
Unfortunately, this simplest variant leads to non-conservation of particles and we have no possibility 
to compare it with standard Langevin approach. To correct this situation we add more complex sources, 
which keep the number of species, and technically can be compared with the effects produced by noices 
presented in Langevin equation. 
 
 This paper is organized as follows. In section \ref{sec:sourcesink} we
formulate general formalism for master equation for annihilation reaction 
with random sources and sinks. We recall basic features of Doi "second quantization" formalism and 
we construct basic field theoretic dynamical  action fuctional. 
 Section \ref{sec:AAsourcesink} is devoted to the analysis of 
 possible situations, which can occur, when we consider different space dimensions and canonical dimensions 
 of master fields.     
 Section
\ref{sec:Conclusion} is devoted to discussion of the results and
concluding remarks.

\section{Random sources and sinks in the master equation}
\label{sec:sourcesink}

We will consider the annihilation reaction $A+A\to\emptyset$ in
a random drift field in a more general setup than in
\cite{Hnatich00}. To this end we introduce random sources and sinks
for the particles, e.g., to maintain a steady state in the system.
In most cases this is carried out by including an additive noise
term in the Langevin equation of the stochastic process. Since our
analysis is based on the master equation, this is not quite
appropriate here. Unfortunately, there is no unique way to introduce
random sources in the master equation corresponding to the random
noise of the mean-field (Langevin) description. We use the simplest
choice, described in detail, e.g., in \cite{vanKampen84}, which is
tantamount to amending the problem by reactions $A\to X$ and $Y\to
A$, where $X$ and $Y$ stand for particle baths of the sink and the
source, respectively. In a homogeneous system these reactions give
rise to the master equations
\begin{eqnarray}
\dee{P(t,n)}{t}&=&\mu_+V\left[P(t,n-1)-P(t,n)\right]
\nonumber\\
&+&\mu_-\left[(n+1)P(t,n+1)-nP(t,n)\right]\ldots
\label{MasterSource}
\end{eqnarray}
where $P(t,n)$ is the probability to find $n$ particles at the
time instant $t$ in the system. The ellipsis in (\ref{MasterSource})
stands for the terms describing the annihilation reaction, diffusion
and advection in the system. In (\ref{MasterSource}) $\mu_+$ and $\mu_-$ are the
reaction constants of the creation and annihilation reactions, respectively. The transition rate
has been chosen proportional to the particle number $n$, which seems the quite natural and
also preserves the empty state as an absorbing state. In the transition rate for creation process
$V$ is the volume of the (for the time being) homogeneous system and will be important in passing to
the continuum limit of the inhomogeneous system. We recall that the master equation (\ref{MasterSource})
gives rise to the reaction-rate equation
\beq
\label{RateSource}
\dee{\langle n\rangle}{t}=\mu_+V-\mu_-\langle n\rangle+\ldots
\eeq
where $\langle n\rangle$ is the mean particle number.

The idea of the Doi approach \cite{Doi76} is to rewrite the set of master
equations for probability distributions of a stochastic problem in
the form of a single kinetic equation for a state vector
incorporating all probabilistic information about the system
constructed in a suitable Fock space. The kinetic equation is
defined by the Liouville operator acting in the Fock
space and generated by the set of master equations. Although the
basic procedure has been thoroughly exposed in the literature, the
introduction of random sources and sinks of particles in the master
equation has specific features which should be presented in detail.
Therefore,
let us briefly recall the basic quantities and relations of the Doi
approach. For simplicity, consider probabilities $P(t,n)$ to find
$n$ particles at the time instant $t$ on a fixed lattice site. The spatial dependence
may then be described by labeling the particle number by the coordinates of the
lattice and introducing necessary sums and products over the lattice sites.

Basis vectors of the Fock space are defined by the usual
annihilation and creation operators $\ann$, $\cre$ and the vacuum
vector $\ket{0}$:
\begin{align}
\label{adefs}
\ann\ket{0}&=0\,,\quad
\cre\ket{n}=\ket{n+1}\,,\\ [\,\ann,\cre]&=\ann\cre-\cre\ann=I\,,
\end{align}
with the normalization $\braket{n}{m}=n!\delta_{nm}$.

The state vector containing all information about the system is defined as
\begin{equation}
\label{statevector} \ket{\Phi}=\sum\limits_{n=0}^\infty
P(t,n)\ket{n}\,.
\end{equation}
The set of master equations for a birth-death process may be cast in the
form of a single evolution equation for the state vector
(\ref{statevector}) without any explicit dependence on the
occupation number:
\begin{equation}
\label{L} \dee{\ket{\Phi}}{t}=\Lio(\cre,\,\ann)\ket{\Phi}\,.
\end{equation}
Master equations (\ref{MasterSource}) give rise to the following terms in
the Liouville operator
\beq
\label{LSource}
\Lio_g(\cre,\ann)= \mu_+V\left(\cre-I\right)+\mu_-\left(I-\cre\right)\ann\,,
\eeq
where $I$ is the identity operator. The expectation value of any function $F(n)$ of the
random particle number
\begin{equation}
\label{average}
\langle F(t)\rangle=\sum\limits_{n=0}^\infty F(n)P(t,n)\,,
\end{equation}
may be expressed in the form of the functional integral over the
functions of time $\tilde{a} $ and $a$:
\begin{equation}
\label{functional integral}
\langle F(t)\rangle=\int\!{\cal D}\tilde{a} {\cal D}a\, F_N[(\tilde{a} (t)+1)a(t)]
e^{S_1}\,,
\end{equation}
where $F_N(\tilde{a} a)$ is the normal form \cite{Vasilev98} of the operator $F(\cre\ann)$ and $S_1$ is the
dynamic action
\begin{equation}
\label{action1}
S_1(\tilde{a} ,a)=\int\limits_0^\infty\!\! dt\,\left[-\tilde{a} (t)\partial_t a(t)
+\mu_+V\tilde{a} (t)-\mu_-\tilde{a} (t)a(t)\right]\ldots
\end{equation}
Only the generic time-derivative term and terms brought about by the random source model are expressed
here explicitly, while the ellipsis stands for terms corresponding to other reactions and initial conditions.

Let the transition rates $\mu_\pm$ be random functions uncorrelated in time with a probability distribution given
in terms of the moments $\langle\mu_\pm^n\rangle=E_{\pm\,,n}$. At this point we also generalize the treatment to the
case of a spatially inhomogeneous system and introduce a lattice subscript as the spatial argument, e.g.
$a(t)\to a_i(t)$. In this case the volume $V$ becomes the volume element attached to the lattice site.
For simplicity, we replace the time integral with the integral sum $\int_0^\infty\! dt\to \sum_\alpha\Delta t$
and assume that the transition rates at each time instant and lattice site $\mu_{\pm\\,\alpha,i}$
are independent random variables. Then the average of the expectation value (\ref{functional integral})
over the distribution of random sources reduces to the calculation of the expectation value
\beq
\label{g-average}
\prod_{\alpha,i}
\langle e^{\mu_{+\,,\alpha,i}V\tilde{a} _{\alpha,i}\Delta t-
\mu_{-\,,\alpha,i}\tilde{a} _{\alpha,i}a_{\alpha,i}\Delta t}\rangle\,.
\eeq
For each particular time instant and lattice this yields (we assume that the moments of $\mu_\pm$ are the same for
all $\alpha$ and $i$ and omit labels for brevity) this gives rise to the usual cumulant expansion
\begin{eqnarray}
\nonumber
\langle e^{\mu b\Delta t}\rangle&=&
1+b\Delta t E_1+{1\over 2}E_2(b\Delta t)^2+{1\over 6}E_3(b\Delta t)^3+\cdots\\
\label{expansion-average}
&=&e^{b\Delta t E_1+{1\over 2}\left(E_2-E_1^2\right)(b\Delta t)^2
+{1\over 6}\left(E_3-3E_1E_2+E_1^3\right)(b\Delta t)^3+\cdots}
\end{eqnarray}
Here, $b$ stands for either $V\tilde{a} $ or $-\tilde{a} a$. Thus, for instance the average over $\mu_+$ assumes the form
\begin{eqnarray}
\nonumber
\prod_{\alpha,i}
\langle e^{\mu_{+\,,\alpha,i}V\tilde{a} _{\alpha,i}\Delta t}\rangle
&=&e^{\sum_\alpha\sum_i\left[\Delta t E_{+1}V\tilde{a} _{\alpha,i}
+{1\over 2}\left(E_{+2}-E_{+1}^2\right)(V\tilde{a} _{\alpha,i}\Delta t)^2\right]}\\
&\times&e^{\sum_\alpha\sum_i\left[{1\over 6}\left(E_{+3}-3E_{+1}E_{+2}+E_{+1}^3\right)(V\tilde{a} _{\alpha,i}\Delta t)^3+\cdots\right]}\,.
\label{+average}
\end{eqnarray}
In the continuum limit the function $\tilde{a} _{\alpha,i}$ is replaced
by the field $\psi^+(t,\vx)$, whereas in the limit $V\to 0$ the expression $a_{\alpha,i}/V$ gives rise to the field
$\psi(t,\vx)$. The sum over $\alpha$ together with $\Delta t$ gives rise to the time integral and the
sum over $i$ together with the volume element gives rise to the spatial integral $\sum_iV\to \int d\vx$.
In the first term of the exponential in (\ref{+average}) we thus obtain
\[
\sum_\alpha\sum_i\Delta t E_{+1}V\tilde{a} _{\alpha,i}\to E_{+1}\int dt\int d\vx\,\psi^+(t,\vx)\,.
\]
In the cumulants of second and higher order the continuum limit is not so obvious.
We assume the simplest nontrivial distribution for $\mu_\pm$, in which only the variance term has
a finite limit, when $\Delta t\to 0$ and $V\to 0$, whereas the contributions of higher-order cumulants
vanish, for instance
\begin{eqnarray}
\label{limit variance}
\left(E_{+2}-E_{+1}^2\right)V\Delta t \to \sigma_+\,,\quad \Delta t\to 0\,,\ V\to 0\,,\\
\label{limit cumulants}
\left(E_{+3}-3E_{+1}E_{+2}+E_{+1}^3\right)(V\Delta t)^2\to 0\,,\quad \Delta t\to 0\,,\ V\to 0\,.
\end{eqnarray}
Therefore, the contribution of the average over $\mu_+$ to the effective dynamic action assumes the form
\beq
\label{+}
S_+= \int\! dt\int\! d\vx\,\left\{E_{+1}\psi^+(t,\vx)+{1\over 2}\sigma_+\left[\psi^+(t,\vx)\right]^2\right\}\,.
\eeq
For the average over $\mu_-$ a similar argument yields
\beq
\label{-}
S_-= \int\! dt\int\! d\vx\,\left\{-E_{-1}\psi^+(t,\vx)\psi(t,\vx)
+{1\over 2}\sigma_-\left[\psi^+(t,\vx)\psi(t,\vx)\right]^2\right\}\,.
\eeq
These contributions to the effective dynamic action may, of course, be generated by suitably chosen normal
distributions of $\mu_\pm$.

This way of introduction of random sources and sinks has the annoying feature that it does not conserve the
number of particles in the system. For a comparison with the treatment of this problem in the Langevin approach the
random sources and sinks should be introduced in such a way that the particle number is conserved.

The simplest way to effect this is to add to the random source a term proportional to the particle number, i.e.
use the ''reaction constant'' $\mu_+V+\mu_{1+}n$ instead of $\mu_+V$ in the master equation. The source terms on
the right-hand side of the master equation (\ref{MasterSource}) in this case the assume the form
\begin{eqnarray}
\nonumber
\dee{P(t,n)}{t}&=&\mu_+V\left[P(t,n-1)-P(t,n)\right]\\
&+&\mu_{1+}\left[(n-1)P(t,n-1)-nP(t,n)\right]\ldots
\label{MasterSource2}
\end{eqnarray}
The new part of the master equation corresponds to a branching process \cite{vanKampen84}.

The added term gives rise to the following contribution to the Liouville operator
\beq
\label{LSource2}
\Lio_{g2}(\cre,\ann)= \mu_{1+}\left(\cre-I\right)\cre\ann\,.
\eeq
Performing the steps described above we arrive at the contribution to the dynamic action in
the form
\beq
\label{1+}
S_{1+}= \int\! dt\int\! d\vx\,\left\{E_{1+1}\psi^+\left(\psi^++1\right)\psi
+{1\over 2}\sigma_{1+}{\psi^+}^2\left(\psi^++1\right)^2\psi^2\right\}\,.
\eeq
It is easy to see now that if we exclude the plain source (i.e. put $E_{+1}=\sigma_+=0$) and choose
$E_{1+1}=E_{-1}$, then the empty state remains absorbing and the ''mass term'' $\propto \psi^+\psi$
disappears in the dynamic action and we arrive at the dynamic action of random sources and sinks
\begin{eqnarray}
\nonumber
S_{gc}&=& \int\! dt\int\! d\vx\,\biggl\{E_{1+1}{\psi^+}^2\psi+
{1\over 2}\sigma_-\left(\psi^+\psi\right)^2\\
&+&{1\over 2}\sigma_{1+}{\psi^+}^2\left(\psi^++1\right)^2\psi^2\biggl\}
\label{gc}
\end{eqnarray}
which conserves the average number of particles.

The effects of the high-order terms are drastically different
in the two cases amenable for a scaling analysis with the aid of the renormalization group.
The time derivative term in the dynamic action
\[
S= -\int\! dt\int\! d\vx\,\psi^+(t,\vx)\partial_t\psi(t,\vx)+\ldots
\]
must be dimensionless in order to have nontrivial dynamics. Therefore
the total scaling dimension of the number-density operator $\psi^+(t,\vx)\psi(t,\vx)$ is equal to
the dimension of space and thus positive.

First, if the
scaling dimension of the field $\psi^+$ is equal to zero, $d_{\psi^+}=0$, then the dimension of the
field $\psi$ is positive (more precisely, $d_{\psi}=d$)
and the operator monomials in the second and third terms in (\ref{gc})
have the same scaling dimension. Since they are carrying the factor $\psi^2$, the their scaling dimension
is larger than that of ${\psi^+}^2\psi$. Therefore, the second and third terms
in (\ref{gc}) are IR irrelevant and should be discarded in the asymptotic analysis. Second,
if the scaling dimensions of both fields are positive, then in the operator monomials
in the second and third terms in (\ref{gc}) there is at least one ''excessive'' field factor
in comparison with the first term which renders them irrelevant.
Thus, in these cases
the IR relevant dynamic action of random sources and sinks reduces to the single term
\beq
\label{gcPrime}
S'_{gc}= \int\! dt\int\! d\vx\,E_{1+1}{\psi^+}^2\psi\,,\quad d_{\psi^+}=0 \ \vee \ d_{\psi^+}>0\,, \ d_{\psi}>0 \,.
\eeq
Third,
if the scaling dimension of the field $\psi=0$, the scaling dimension
of the field $\psi^+$ is positive, terms with ''excessive'' powers of $\psi^+$ are
IR irrelevant and the starting point for the subsequent RG analysis is the source and sink action
in the form
\beq
\label{gcPos}
S_{gc}''= \int\! dt\int\! d\vx\,\left\{E_{1+1}{\psi^+}^2\psi+
{1\over 2}\left(\sigma_-+\sigma_{1+}\right)\left(\psi^+\psi\right)^2\right\}\,,
\quad d_{\psi}=0\,.
\eeq

\section{Annihilation reaction $A+A\to\emptyset$ with random sources and sinks}
\label{sec:AAsourcesink}

Let us now analyze the dynamic action of the diffusion-limited annihilation reaction
$A+A\to\emptyset$
\begin{eqnarray}
\nonumber
S_1&=&-\int_0^\infty\!\! dt\!\int\! d{\bf x} \,\biggl\{\psi^+\partial_t\psi
-D_0 \psi^+\nabla^2\psi\\
&+&\lambda_0D_0\left[2\psi^++
(\psi^+)^2\right]\psi^2
\biggl\}
+n_0\int\! d{\bf x}\, \psi^+({\bf x},0)\,.
\label{action11}
\end{eqnarray}
from the point of view of scaling behaviour sketched in section \ref{sec:sourcesink}.

In the first case with $d_{\psi^+}=0$ the nonlinear terms in action (\ref{action11})
are of equal scaling dimension. However, the source-sink part (\ref{gcPrime}) is linear
in the field $\psi$ with positive scaling dimension in contrast to the quadratic
in $\psi$ terms of (\ref{action11}). Therefore, the IR relevant interaction term in this
case is (\ref{gcPrime}) and the corresponding dynamic action is
\begin{eqnarray}
\nonumber
S_{IR1}&=&-\int_0^\infty\!\! dt\!\int\! d{\bf x} \,\left\{\psi^+\partial_t\psi
-D_0 \psi^+\nabla^2\psi
-E_{1+1}{\psi^+}^2\psi
\right\}\\
&+&n_0\int\! d{\bf x}\, \psi^+({\bf x},0)\,.
\label{IRaction1}
\end{eqnarray}
However, this dynamic action does not bring about any graphs with closed loops of the
density propagator and thus no density fluctuation effects on the asymptotic behaviour are
anticipated.

In the second case with $d_{\psi^+}>0$ and $d_{\psi}>0$ the fourth-order term in action (\ref{action11})
becomes irrelevant. Either of the remaining third-order terms alone does not generate loops, therefore
density fluctuation effects are brought about only, when both fields have the same scaling dimension
$d_{\psi^+}=d_{\psi}=d/2$. In this case the IR relevant dynamic action is
\begin{eqnarray}
\nonumber
S_{IR2}&=&-\int_0^\infty\!\! dt\!\int\! d{\bf x} \,\biggl\{\psi^+\partial_t\psi
-D_0 \psi^+\nabla^2\psi\\
&+&2\lambda_0D_0\psi^+\psi^2 -E_{1+1}{\psi^+}^2\psi
\biggl\}
+n_0\int\! d{\bf x}\, \psi^+({\bf x},0)\,.
\label{IRaction2}
\end{eqnarray}
This is the dynamic action of the Gribov process \cite{Gribov67}, also known as the Reggeon model,
subject to random advection.
Effects of the random drift with the use of the Obukhov-Kraichnan compressible velocity field
have been analyzed in \cite{Antonov10}.

In the third case with $d_{\psi}=0$ the fourth-order term in action (\ref{action11})
becomes irrelevant as well due to the positive dimension of the field $\psi^+$. By the same token, however,
both terms of the source-sink action (\ref{gcPos}) are also irrelevant and we arrive at the IR relevant
dynamic action
\begin{eqnarray}
\nonumber
S_{IR3}&=&-\int_0^\infty\!\! dt\!\int\! d{\bf x} \,\biggl\{\psi^+\partial_t\psi
-D_0 \psi^+\nabla^2\psi
+2\lambda_0D_0\psi^+\psi^2
\biggl\}\\
&+&n_0\int\! d{\bf x}\, \psi^+({\bf x},0)\,.
\label{IRaction3}
\end{eqnarray}
This dynamic action does not give rise to any density-fluctuation loops and thus does not predict any
decay anomaly due to them.

In summary, if the sources and sinks are chosen such that they conserve the mean number of particles in
the system, then the anomalous scaling behaviour in the system is that of the Gribov process, if any.

A different situation arises, if the plain source term is included in the analysis. Then there is the possibility
that the system does not tend to the absorbing empty state but to an active state with a finite concentration
of particles. In this case the starting point is the dynamic action with all the terms quoted above, i.e.
\begin{eqnarray}
\nonumber
S&=&\int_0^\infty\!\! dt\!\int\! d{\bf x} \,\Biggl\{-\psi^+\partial_t\psi
+D_0 \psi^+\nabla^2\psi
-\lambda_0D_0\left[2\psi^++
\left(\psi^+\right)^2\right]\psi^2
+E_{+1}\psi^+\\
\nonumber
&+&{1\over 2}\sigma_+\left(\psi^+\right)^2
+E_{1+1}\psi^+\left(\psi^++1\right)\psi
+{1\over 2}\sigma_{1+}{\psi^+}^2\left(\psi^++1\right)^2\psi^2
-E_{-1}\psi^+\psi\\
\label{actionfull}
&+&{1\over 2}\sigma_-\left(\psi^+\psi\right)^2
\Biggl\}
+n_0\int\! d{\bf x}\, \psi^+({\bf x},0)\,.
\end{eqnarray}
The stationarity equation brought about by this dynamic action for the field $\psi$ is (the stationary
value $\psi^+=0$ as usual)
\beq
\label{statfull}
\partial_t\psi
-D_0 \nabla^2\psi=-2\lambda_0D_0\psi^2
+E_{+1}
+E_{1+1}\psi
-E_{-1}\psi\,.
\eeq
However, the action expanded around the stationary value brought about by this equation is rather complicated.
To keep expressions simple, continue to consider the case $E_{1+1}=E_{-1}$. Then the
re-expanded action is
\begin{eqnarray}
\nonumber
S&=&\int_0^\infty\!\! dt\!\int\! d{\bf x} \Bigl\{-\psi^+\partial_t\psi
+D_0 \psi^+\nabla^2\psi        -{\sqrt{8}}{\sqrt{ E_{+1} {\lambda_0D_0} }} \psi^+ \psi \\
 \nonumber
 &+&\left( \frac{- E_{+1} }{2} +
       \frac{ E_{-1} {\sqrt{ E_{+1} \,{\lambda_0D_0} }}}{{\sqrt{2}}{\lambda_0D_0} } +
       \frac{ E_{+1} \sigma_{1+} }{4{\lambda_0D_0} } +
       \frac{ E_{+1}  \sigma_{-} }{4 {\lambda_0D_0} } + \frac{ \sigma_+}{2} \right)
      { \psi^+ }^2\\
\nonumber
&+& \frac{ E_{+1}  \sigma_{1+} { \psi^+ }^3}
     {2{\lambda_0D_0} } + \frac{ E_{+1}  \sigma_{1+} { \psi^+ }^4}{4{\lambda_0D_0} }
 + \frac{{\sqrt{2}}\,{\sqrt{ E_{+1} \,{\lambda_0D_0} }}\,
         \sigma_{1+} \,{ \psi^+ }^3\psi}{{\lambda_0D_0} } \\
     \nonumber
&+&\left(  E_{-1}  - {\sqrt{2}}\,{\sqrt{ E_{+1} \,{\lambda_0D_0} }} +
        \frac{{\sqrt{ E_{+1} \,{\lambda_0D_0} }}\, \sigma_{1+} }{{\sqrt{2}}\,{\lambda_0D_0} } +
        \frac{{\sqrt{ E_{+1} \,{\lambda_0D_0} }}\, \sigma_{-} }{{\sqrt{2}}\,{\lambda_0D_0} } \right){ \psi^+ }^2\psi
      \\
\nonumber      
&+&
     \frac{{\sqrt{ E_{+1} \,{\lambda_0D_0} }}\, \sigma_{1+} \,{ \psi^+ }^4\psi}
      {{\sqrt{2}}\,{\lambda_0D_0} }
        -2\,{\lambda_0D_0} \, \psi^+{\psi }^2  +
     \left( -{\lambda_0D_0}  + \frac{ \sigma_{1+} }{2} + \frac{ \sigma_{-} }{2} \right) \,
      { \psi^+ }^2 {\psi }^2\\
&+&  \sigma_{1+} \,{ \psi^+ }^3 {\psi }^2+
     \frac{ \sigma_{1+} \,{ \psi^+ }^4{\psi }^2}{2}
     \Bigr\}
+n_0\int\! d{\bf x}\, \psi^+({\bf x},0)\,.
\label{actioneffective}
\end{eqnarray}
In the critical limit $E_{+1}\to 0$. Since it is the expectation value of a nonnegative random
quantity $\mu_+$, the variance $\sigma_+$ vanishes as well. In the vicinity of the critical point
we keep only the leading $E_{+1}$ and $\sigma_+$ putting them equal to zero in terms, where they are
subleading. This simplifies the action a bit

\begin{eqnarray}
\nonumber
S&=&\int_0^\infty\!\! dt\!\int\! d{\bf x}\Bigl\{-\psi^+\partial_t\psi
+D_0 \psi^+\nabla^2\psi        -{\sqrt{8}}{\sqrt{ E_{+1}{\lambda_0D_0} }} \psi^+ \psi \\
 \nonumber
 &+&\left(
       \frac{ E_{-1} \,{\sqrt{ E_{+1}  }}}{{\sqrt{2{\lambda_0D_0}}}\, } + \frac{ \sigma_+}{2} \right)
      \,{ \psi^+ }^2
      + \frac{ E_{+1} \, \sigma_{1+} \,{ \psi^+ }^3}
     {2\,{\lambda_0D_0} } + \frac{ E_{+1} \, \sigma_{1+} \,{ \psi^+ }^4}{4\,{\lambda_0D_0} }
+  E_{-1}  \,
      { \psi^+ }^2\psi\\
\nonumber
&+& \frac{{\sqrt{2}}\,{\sqrt{ E_{+1} \,{\lambda_0D_0} }}\,
         \sigma_{1+} \,{ \psi^+ }^3\psi}{{\lambda_0D_0} }
          +
     \frac{{\sqrt{ E_{+1} \,{\lambda_0D_0} }}\, \sigma_{1+} \,{ \psi^+ }^4\psi}
      {{\sqrt{2}}\,{\lambda_0D_0} }
        -2\,{\lambda_0D_0} \, \psi^+{\psi }^2 \\ 
 \nonumber
 &+&
     \left( -{\lambda_0D_0}  + \frac{ \sigma_{1+} }{2} + \frac{ \sigma_{-} }{2} \right) \,
      { \psi^+ }^2 {\psi }^2+  \sigma_{1+} \,{ \psi^+ }^3 {\psi }^2+
     \frac{ \sigma_{1+} \,{ \psi^+ }^4{\psi }^2}{2}
     \Bigr\}\\
&+&n_0\int\! d{\bf x}\, \psi^+({\bf x},0)\,.
\label{actioneffective2}
\end{eqnarray}
Dimensional analysis of the canonical dimensions then yields the following cases. In the nonlinear parts
without the critical parameters  $E_{+1}$ and $\sigma_+$  the
previous arguments hold, but in terms having powers of these parameters as coefficients the positive scaling
dimensions of them must be taken into account. The free-field part of the action (\ref{actioneffective2})
suggests that the canonical dimension of $E_{+1}$ is four. The canonical dimension of $\sigma_+$, in fact, remains
a free parameter.

Proceeding in the same manner as above, we arrive at the following effective actions for the IR scaling limit.
In the first case with $d_{\psi^+}=0$ the third and fourth powers of ${\psi^+}$ and independent of ${\psi}$
or first order in ${\psi}$
due to the coefficients proportional to $E_{+1}$ or its square root are irrelevant compared with terms $\propto{\psi^+}^2$
in action (\ref{actioneffective2}). Nonlinear in ${\psi}$ terms are irrelevant against the linear terms due to
positive dimension of ${\psi}$.
Therefore, the IR effective action in this case is
\begin{eqnarray}
\nonumber
S&=&\int_0^\infty\!\! dt\!\int\! d{\bf x} \,\Bigl\{-\psi^+\partial_t\psi
-\psi^+\nabla [{\bf v}\psi]
+D_0 \psi^+\nabla^2\psi        -2\,{\sqrt{2}}\,{\sqrt{ E_{+1} \,{\lambda_0D_0} }}\, \psi^+ \psi\\
 &+&\left(
       \frac{ E_{-1} \,{\sqrt{ E_{+1}  }}}{{\sqrt{2{\lambda_0D_0}}}\, } + \frac{ \sigma_+}{2} \right)
      \,{ \psi^+ }^2
      +  E_{-1}  \,
      { \psi^+ }^2\psi
     \Bigr\}
+n_0\int\! d{\bf x}\, \psi^+({\bf x},0)\,.
\label{IRactionE+1}
\end{eqnarray}
Again, the only remaining interaction terms does not bring about loops, although here we have a nontrivial
correlation function of the field ${\psi}$.

In the second case with $d_{\psi^+}>0$ and $d_{\psi}>0$ higher powers than the leading corrections
to the free-field action of both fields are irrelevant. This argument leaves us with the dynamic action
\begin{eqnarray}
\nonumber
S&=&\int_0^\infty\!\! dt\!\int\! d{\bf x} \,\Bigl\{-\psi^+\partial_t\psi
-\psi^+\nabla [{\bf v}\psi]
+D_0 \psi^+\nabla^2\psi        -2\,{\sqrt{2}}\,{\sqrt{ E_{+1} \,{\lambda_0D_0} }}\, \psi^+ \psi\\
\nonumber
 &+&\left(
       \frac{ E_{-1} \,{\sqrt{ E_{+1}  }}}{{\sqrt{2{\lambda_0D_0}}}\, } + \frac{ \sigma_+}{2} \right)
      \,{ \psi^+ }^2
      +  E_{-1}  \,
      { \psi^+ }^2\psi   -2\,{\lambda_0D_0} \, \psi^+{\psi }^2
     \Bigr\}\\
&+&n_0\int\! d{\bf x}\, \psi^+({\bf x},0)\,.
\label{IRactionE+2}
\end{eqnarray}
Contrary to the case discussed above, here the interaction term $-2\,{\lambda_0D_0} \, \psi^+{\psi }^2$
generates loops alone due to the presence of the correlation function of the field ${\psi}$. Therefore,
two effective actions with nontrivial fluctuation contributions are possible.

a)
$d_{\psi^+}>d_{\psi}$. To keep the correlation function of the field ${\psi}$ for the loops, the variance
$\sigma_+$ must have a dimension less than that of $\sqrt{E_{+1}}$. This yields the effective action
\begin{eqnarray}
\nonumber
S&=&\int_0^\infty\!\! dt\!\int\! d{\bf x} \,\Bigl\{-\psi^+\partial_t\psi
-\psi^+\nabla [{\bf v}\psi]
+D_0 \psi^+\nabla^2\psi        -2\,{\sqrt{2}}\,{\sqrt{ E_{+1} \,{\lambda_0D_0} }}\, \psi^+ \psi\\
 &+&\frac{ \sigma_+}{2}
      \,{ \psi^+ }^2
  -2\,{\lambda_0D_0} \, \psi^+{\psi }^2
     \Bigr\}
+n_0\int\! d{\bf x}\, \psi^+({\bf x},0)
\label{IRactionE+2a}
\end{eqnarray}
with the critical dimension depending on the scaling dimension of $\sigma_+$ in the spirit of the
description of tricritical scaling behaviour \cite{Vasiljev}. The model is logarithmic at six dimensions, however,
because apart from the coefficient of the $\propto {\psi^+}^2$ the action is that of critical dynamics of the
$\varphi^3$ model.

b) $d_{\psi^+}=d_{\psi}=d/2$. Both third-order terms are relevant and the effective action is
basically (\ref{IRactionE+2}). In this case the dimension of
$\sigma_+$ is larger than that of $\sqrt{E_{+1}}$ and for simplicity we omit $\sigma_+$. Thus, the
effective dynamic action may be written as
\begin{eqnarray}
\nonumber
S&=&\int_0^\infty\!\! dt\!\int\! d{\bf x} \,\Bigl\{-\psi^+\partial_t\psi
-\psi^+\nabla [{\bf v}\psi]
+D_0 \psi^+\nabla^2\psi        -2\,{\sqrt{2}}\,{\sqrt{ E_{+1} \,{\lambda_0D_0} }}\, \psi^+ \psi\\
 &+&
       \frac{ E_{-1} \,{\sqrt{ E_{+1}  }}}{{\sqrt{2{\lambda_0D_0}}}\, }
      \,{ \psi^+ }^2
      +  E_{-1}  \,
      { \psi^+ }^2\psi   -2\,{\lambda_0D_0} \, \psi^+{\psi }^2
     \Bigr\}
+n_0\int\! d{\bf x}\, \psi^+({\bf x},0)\,.
\label{IRactionE+2b}
\end{eqnarray}
Note that this is a dynamic action describing the Gribov process with a random source independent of
the active agent density. That the rate of change of the density due to the random sink is proportional
to a power of density is a natural assumption. The assumption that the rate of change of the density due
to the random source is proportional to a power of density is not natural. Therefore, the dynamic action
(\ref{IRactionE+2b}) possibly predicts a critical behaviour of the Gribov process different from that
discussed in the literature.

In the third case with $d_{\psi^+}>0$ and $d_{\psi}=0$ we arrive at the effective action (\ref{IRactionE+2a}).

The analysis of scaling dimensions shows that we may actually lift most of the restrictions on the probability
distribution of the transition rates of the type (\ref{limit variance}) and (\ref{limit cumulants}).
Indeed, even if the higher order cumulants are finite, the scaling dimensions of corresponding terms
in the dynamic action grow with the order of the cumulant with the exception of the case, when the transition
rate is independent of the agent density.



\section{Conclusion}
\label{sec:Conclusion}
We have investigated possible effects of random sources and sinks on annihilation reaction. Contrary to frequently used approach, in which  the sources and sinks are introduced directly to the Langevin equation, we have added it directly to the master equation, where their physical sense is clear. 
On the basis of dimensional analysis we have constructed effective actions, which are a starting point for RG analysis of critical behaviour of systems under consideration.
Most of the dynamic effective actions obtained are trivial from the point of view of critical behaviour of the model, because infrared relevant interaction terms do not generate loops diagrams. We can conclude that for the study of influence of density fluctuations on annihilation reaction $A+A\to\emptyset$, which are relevant only for dimension two or less, random sources and sinks are unimportant. 
Rest non-trivial cases lead to the action describing Gribov process. However, in this manner we obtain a source term in dynamic action without agent density. This is cardinally different from that generated by Langevin equation, therefore RG analysis can lead to the new
effects in critical  behaviour, when sources and sinks asymptotically vanish. We need analyse dependence of scaling functions on sinks and sources parameters in infrared limit. It suggests the situation, which takes place in phase transitions, where statistical correlations of order parameter depends on "mass" (deviation of temperature from Currie one) and dependence of their scaling functions on "`mass" have to be investigated. 
This analysis we intend to carry out in near future.\\
 The work was supported by VEGA grant 0173 of Slovak Academy of Sciences, and by Centre of Excellency for
Nanofluid of IEP SAS. This article was also created by implementation
of the "Cooperative phenomena and phase transitions
in nanosystems with perspective utilization in nano- and
biotechnology" projects No 26220120021 and No 26220120033. Funding for the operational
research and development program was provided by
the European Regional Development Fund.

\end{document}